\documentclass[prd,amsfonts,noshowpacs,nofootinbib]{revtex4}
\usepackage{amsmath}

\begin{document}

\begin{abstract}
We study the non-gaussianity from the bispectrum in multi-field
inflation models with a general kinetic term. The models include
the multi-field K-inflation and the multi-field Dirac-Born-Infeld
(DBI) inflation as special cases. We find that, in general, the
sound speeds for the adiabatic and entropy perturbations are
different and they can be smaller than 1. Then the non-gaussianity
can be enhanced. The multi-field DBI-inflation is shown to be a
special case where both sound speeds are the same due to a special
form of the kinetic term. We derive the exact second and third
order actions including metric perturbations. In the small sound
speed limit and at leading order in the slow-roll expansion, we
derive the three point function for the curvature perturbation
which depends on both adiabatic and entropy perturbations. The
contribution from the entropy perturbations has a different
momentum dependence if the sound speed for the entropy
perturbations is different from the adiabatic one, which provides
a possibility to distinguish the multi-field models from single
field models. On the other hand, in the multi-field DBI case, the
contribution from the entropy perturbations has the same momentum
dependence as the pure adiabatic contributions and it only changes
the amplitude of the three point function. This could help to ease
the constraints on the DBI-inflation models.
\end{abstract}

\title{Non-gaussianity from the bispectrum in general multiple field inflation}
\author{Frederico Arroja $\dagger$\footnote{Frederico.Arroja@port.ac.uk}}
\author{Shuntaro Mizuno
$\ddagger$\footnote{mizuno@resceu.s.u-tokyo.ac.jp}}
\author{Kazuya Koyama $\dagger$ \footnote{Kazuya.Koyama@port.ac.uk}}
\affiliation{$\dagger$Institute of Cosmology and Gravitation, University of
Portsmouth, Portsmouth PO1 2EG, UK \\
$\ddagger$ Research Center for the Early Universe (RESCEU), School of
Science, University of Tokyo, 7-3-1 Hongo, Bunkyo, Tokyo~113-0033,
Japan
}
\date{\today}
\maketitle

%%%%%%%%%%%%%%%%%%%%%%%%%%%%%%%%%%%%%%%%%%%%%%%%%%%%%%%%%%%%%%%%%%%%%%%%%%%%%%%%%%%%%%%%%%%%%%%%%%%%%%%%%%%%%%%%%%%%%%%%%%%
%%%%%%%%%%%%%%%%%%%%%%%%%%%%%%%%%%%%%%%%%%%%%%%%%%%%%%%%%%%%%%%%%%%%%%%%%%%%%%%%%%%%%%%%%%%%%%%%%%%%%%%%%%%%%%%%%%%%%%%%%%%
\section{Introduction}
The inflationary scenario succeeds to explain the origin of
temperature fluctuations of the Cosmic Microwave Background (CMB).
The increasing precision of the measurements of the CMB enables us
to distinguish between many inflationary models. The primordial
fluctuations generated during inflation are nearly scale invariant
and gaussian. Thus the deviation from the exact scale invariance
and gaussianity will give valuable information in discriminating
many possible models. Especially, non-gaussianity of the
primordial fluctuations will provide powerful ways to constrain
models (see e.g. \cite{Bartolo:2004if} for a review). The simplest
single field inflation models predict that the non-gaussianity of
the fluctuations should be very difficult to be detected even in
the future experiments such as Planck \cite{Planck}. If we detect
large non-gaussianity, this means that the simplest model of
inflation would be rejected.

There are a few models where the primordial fluctuations generated
during inflation have a large non-gaussianity. In the single field
case, if the inflaton field has a non-trivial kinetic term, it is
known that the non-gaussianity can be large. In $K$-inflation
models where the kinetic term of the inflaton field is generic,
the sound speed of the perturbations can be much smaller than $1$
\cite{ArmendarizPicon:1999rj,Garriga:1999vw}. This leads to a
large non-gaussianity of the fluctuations. The Dirac-Born-Infeld
(DBI) inflation, motivated by string theory, is another example
\cite{Silverstein:2003hf,Alishahiha:2004eh,Chen:2004gc,Chen:2005ad}.
In DBI-inflation, the inflaton is identified with the position of
a moving D3 brane whose dynamics is described by the DBI action.
Again, due to the non-trivial form of the kinetic term, the sound
speed can be smaller than $1$ and the non-gaussianity becomes
large \cite{Chen:2005fe,Chen:2006nt}. The third and fourth order
actions for a single inflaton field with a generic kinetic term
have been calculated by properly taking into account metric
perturbations and three and four point functions have been
calculated \cite{Chen:2006nt,Huang:2006eh,Arroja:2008ga}. For the
detailed observational consequences of single-field DBI-inflation
see
Refs.~\cite{Kecskemeti:2006cg,Lidsey:2006ia,Baumann:2006cd,Bean:2007hc,
Lidsey:2007gq,Peiris:2007gz,Kobayashi:2007hm,Lorenz:2007ze}.

Multi-field inflation models where the curvature perturbation is
modified on large scales due to the entropy perturbations
\cite{Starobinsky:1994mh} have been also extensively studied
recently. In the case of the standard kinetic term, it is not easy
to generate large non-gaussianity from multi-field dynamics
\cite{Rigopoulos:2005ae, Vernizzi:2006ve} (see however
Ref.~\cite{Sasaki:2008uc}). In the DBI-inflation case, the
position of the brane in each compact direction is described by a
scalar field. Then DBI-inflation is naturally a multi-field model
\cite{Easson:2007dh}. The effect of the entropy perturbations in
the inflationary models based on string theory constructions in a
slightly different context is also considered in
\cite{Lalak:2007vi,Brandenberger:2007ca}. Recently, Huang
\emph{et.al.} calculated the bispectrum of the perturbations in
multi-field DBI-inflation {\it with the assumption that} the
kinetic term depends only on $X=-G^{IJ} \partial_{\mu} \phi^I
\partial^{\mu} \phi^J /2$ where $\phi^J$ are the scalar fields
$(I=1,2,...)$ and $G_{IJ}$ is the metric in the field space as in
the case of K-inflation \cite{Huang:2007hh}. They found that in
addition to the usual bispectrum of adiabatic perturbations, there
exists a new contribution coming from the entropy perturbations.
Then they showed that the entropy field perturbations propagate
with the speed of light and the contribution from the entropy
perturbations is suppressed. This property can also be confirmed
by the analysis of a more general class of multi-field models
where the kinetic terms are given by arbitrary functions of X
\cite{Langlois:2008mn,Gao:2008dt}. However, Langlois \emph{et.al.}
pointed out that their assumption {\it cannot} be justified for
the multi-field DBI-inflation \cite{Langlois:2008wt}. Even though
the action depends only on $X$ in the homogeneous background,
there exit other kind of terms which contribute only to
inhomogeneous perturbations. They find that this dramatically
changes the behavior of the entropy perturbations. In fact, it was
shown that the entropy perturbations propagate with the same sound
speed as the adiabatic perturbations.

In this paper, we study a fairly general class of multi-field
inflation models with a general kinetic term which includes
K-inflation and DBI-inflation. We study the sound speeds of the
adiabatic perturbations and entropy perturbations and clarify the
difference between K-inflation and DBI-inflation. Then we
calculate the third order action by properly taking into account
the effect of gravity. Then three point functions at leading
order in slow-roll and in the small sound speed limit are
obtained. We can recover the results for K-inflation and
DBI-inflation easily from this general result.

The structure of the paper is as follows. In section II, we
describe our model and derive the equations in the background. In
section III, we study the perturbations using the ADM formalism.
The second and third order actions are derived by properly taking
into account the metric perturbations. Then we decompose the
perturbations into adiabatic and entropy directions and write down
the action in terms of the decomposed fields. In section IV, we
study the sound speed in several models including K-inflation and
DBI-inflation. It is shown that in general, adiabatic and entropy
sound speeds are different and both can be smaller than 1. It is
shown that the DBI action can be obtained by requiring that the
sound speeds for the adiabatic and entropy perturbations are the
same. In section V, the third order action at leading order in
slow-roll and in the small sound speed limit is obtained in terms
of the decomposed fields. Then the three point functions are
derived for a generalized model which includes K-inflation and
DBI-inflation as particular cases. Section VI is devoted to the
conclusion.

%%%%%%%%%%%%%%%%%%%%%%%%%%%%%%%%%%%%%%%%%%%%%%%%%%%%%%%%%%%%%%%%%%%%%%%%%%%%%%%%%%%%%%%%%%%%%%%%%%%%%%%%%%%%%%%%%%%%%%%%%%%
%%%%%%%%%%%%%%%%%%%%%%%%%%%%%%%%%%%%%%%%%%%%%%%%%%%%%%%%%%%%%%%%%%%%%%%%%%%%%%%%%%%%%%%%%%%%%%%%%%%%%%%%%%%%%%%%%%%%%%%%%%%
\section{\label{sec:MODEL}The model}
We consider a very general class of models described by the following action
\begin{equation}
S=\frac{1}{2}\int
d^4x\sqrt{-g}\left[M^2_{Pl}R+2P(X^{IJ},\phi^I)\right],
\label{action}
\end{equation}
where $\phi^I$ are the scalar fields $(I=1,2,...,N)$, $M_{Pl}$ is the Planck
mass that we will set to unity hereafter, $R$  is the Ricci scalar
and
\begin{equation}
X^{IJ}\equiv-\frac{1}{2}g^{\mu\nu}\partial_\mu\phi^I\partial_\nu\phi^J,
\end{equation}
is the kinetic term, $g_{\mu\nu}$ is the metric tensor. We label
the fields' Lagrangian by $P$ and we assume that it is a well
behaved function. Greek indices run from 0 to 3. Lower case Latin
letters $(i,j,...)$ denote spatial indices. Upper case Latin letters denote
field indices.

The Einstein field equations in this model are
\begin{equation}
G_{\mu\nu}=Pg_{\mu\nu}+P_{,X^{IJ}}\partial_\mu\phi^I\partial_\nu\phi^J\equiv
T_{\mu\nu},
\end{equation}
where $P_{,X^{IJ}}$ denotes the derivative of $P$ with respect to
$X^{IJ}$ \cite{derivative}. The generalized Klein-Gordon equation reads
\begin{equation}
g^{\mu\nu}\left(P_{,X^{IJ}}\partial_\nu\phi^I\right)_{;\mu}+P_{,J}=0,
\end{equation}
where $;$ denotes covariant derivative with respect to
$g_{\mu\nu}$ and $P_{,J}$ denotes the derivative of $P$ with
respect to $\phi^J$.

In the background, we are interested in flat, homogeneous and
isotropic Friedmann-Robertson-Walker universes described by the
line element
\begin{equation}
ds^2=-dt^2+a^2(t)\delta_{ij}dx^idx^j, \label{FRW}
\end{equation}
where $a(t)$ is the scale factor. The Friedmann equation and the
continuity equation read
\begin{equation}
3H^2=E_0, \label{EinsteinEq}
\end{equation}
\begin{equation}
\dot{E_0}=-3H\left(E_0+P_0\right), \label{continuity}
\end{equation}
where the Hubble rate is $H=\dot{a}/a$, $E_0$ is the total energy
of the fields and it is given by
\begin{equation}
E_0=2X_0^{IJ}P_{,X^{IJ}}-P_0 ,\label{energy}
\end{equation}
where a subscript zero denotes background quantities and
$X_0^{IJ}=1/2\dot \phi_0^I\dot \phi_0^J$. The equations of motion
for the scalar fields reduce to
\begin{equation}
P_{,X^{IJ}}\ddot\phi^I+\left(3HP_{,X^{IJ}}+\dot
P_{,X^{IJ}}\right)\dot \phi^I - P_{,J}=0.\label{KG1}
\end{equation}

%%%%%%%%%%%%%%%%%%%%%%%%%%%%%%%%%%%%%%%%%%%%%%%%%%%%%%%%%%%%%%%%%%%%%%%%%%%%%%%%%%%%%%%%%%%%%%%%%%%%%%%%%%%%%%%%%%%%%%%%%%%
%%%%%%%%%%%%%%%%%%%%%%%%%%%%%%%%%%%%%%%%%%%%%%%%%%%%%%%%%%%%%%%%%%%%%%%%%%%%%%%%%%%%%%%%%%%%%%%%%%%%%%%%%%%%%%%%%%%%%%%%%%%
\section{\label{sec:Perturbations}Perturbations}
In this section, we will consider perturbations of the background
(\ref{FRW}) beyond linear order. For this purpose, we will
construct the action at second and third order in the
perturbations and it is convenient to use the ADM metric formalism
\cite{Chen:2006nt,Arnowitt:1960es,Maldacena:2002vr,Seery:2005wm,Seery:2005gb,
Langlois:2008mn,Gao:2008dt}. The ADM line element
reads
\begin{equation}
ds^2=-N^2dt^2+h_{ij}\left(dx^i+N^idt\right)\left(dx^j+N^jdt\right),
\label{ADMmetricphi}
\end{equation}
where $N$ is the lapse function, $N^i$ is the shift vector and
$h_{ij}$ is the 3D metric.
The action (\ref{action}) becomes
\begin{equation}
S=\frac{1}{2}\int
dtd^3x\sqrt{h}N\left({}^{(3)}\!R+2P(X^{IJ},\phi^I)\right)+
\frac{1}{2}\int dtd^3x\sqrt{h}N^{-1}\left(E_{ij}E^{ij}-E^2\right).
\end{equation}
The tensor $E_{ij}$ is defined as
\begin{equation}
E_{ij}=\frac{1}{2}\left(\dot{h}_{ij}-\nabla_iN_j-\nabla_jN_i\right),
\end{equation}
and it is related to the extrinsic curvature by
$K_{ij}=N^{-1}E_{ij}$. $\nabla_i$ is the covariant derivative with
respect to $h_{ij}$.
$X^{IJ}$ can be written as
\begin{equation}
X^{IJ}=-\frac{1}{2}h^{ij}\partial_i\phi^I\partial_j\phi^J+\frac{N^{-2}}{2}v^Iv^J\,,
\end{equation}
where $v^I$ is defined as
\begin{equation}
v^I\equiv \dot{\phi}^I-N^j\nabla_j\phi^I.\label{vi}
\end{equation}
The Hamiltonian and momentum constraints are respectively
\begin{eqnarray}
{}^{(3)}\!R+2P-2N^{-2}P_{,X^{IJ}}v^Iv^J-N^{-2}\left(E_{ij}E^{ij}-E^2\right)&=&0,\nonumber\\
\nabla_j\left(N^{-1}E_i^j\right)-\nabla_i\left(N^{-1}E\right)&=&
N^{-1}P_{,X^{IJ}}v^I\nabla_i\phi^J.\label{LMphi}
\end{eqnarray}
We decompose the shift vector $N^i$ into scalar and intrinsic
vector parts as
\begin{equation}
N_i=\tilde{N_i}+\partial_i\psi,
\end{equation}
where $\partial_i\tilde{N^i}=0$, here and in the rest of the
section indices are raised with $\delta_{ij}$.

%%%%%%%%%%%%%%%%%%%%%%%%%%%%%%%%%%%%%%%%%%%%%%%%%%%%%%%%%%%%%%%%%%%%%%%%%%%%%%%%%%%%%%%%%%%%%%%%%%%%%%%%%%%%%%%%%%%%%%%%%%%
\subsection{\label{subsec:PerturbationsUniformeR}Perturbations in the uniform curvature gauge}
In the uniform curvature gauge, the 3D metric takes the form
\begin{eqnarray}
&&h_{ij}=a^2\delta_{ij},\nonumber\\
&&\phi^I(x,t)=\phi_0^I(t)+Q^I(x,t),
\label{deltaphigauge}
\end{eqnarray}
where $Q^I$ denotes the field perturbations. In the following, we
will usually drop the subscript $``0"$ on $\phi_0^I$ and simply
identify $\phi^I$ as the homogeneous background fields unless
otherwise stated.

We expand $N$ and $N^i$ in powers of the perturbation $Q^I$
\begin{eqnarray}
N&=&1+\alpha_1+\alpha_2+\cdots,\\
\tilde{N_i}&=&\tilde{N_i}^{(1)}+\tilde{N_i}^{(2)}+\cdots,\\
\psi&=&\psi_1+\psi_2+\cdots,
\end{eqnarray}
where $\alpha_n$, $\tilde{N_i}^{(n)}$ and $\psi_n$ are of order
$(Q^I)^n$. At first order in $Q^I$, a particular solution for
equations (\ref{LMphi}) is:
\begin{eqnarray}
\alpha_1&=&\frac{P_{,X^{IJ}}}{2H} \dot{\phi^I} Q^J, \quad
\tilde{N_i}^{(1)}=0, \nonumber\\
\partial^2\psi_1&=&\frac{a^2}{2H} \left[
-6H^2\alpha_1+P_{,K}Q^K-
\dot{\phi}^I \dot{\phi}^J P_{,X^{IJ}K}Q^K
+\left(P_{,X^{IJ}}+
\dot{\phi}^L \dot{\phi}^M P_{,X^{LM}X^{IJ}}\right)
\left(\dot{\phi}^I \dot{\phi}^J \alpha_1-\dot
\phi^J\dot Q^I\right) \right]\,.
\end{eqnarray}

The second order action is calculated as
\begin{eqnarray}
S_{(2)}=\int
dtd^3x\frac{a^3}{2}&\Bigg[&X_1^{IJ}X_1^{LM}P_{,X^{IJ}X^{LM}}
+P_{,X^{IJ}}\dot Q^I\dot Q^J - a^{-2}P_{,X^{IJ}}\partial^i
Q^I\partial_iQ^J -\frac{3}{2}\dot
\phi^J\dot \phi^LP_{,X^{IJ}}P_{,X^{LM}}Q^IQ^M \nonumber\\
&&+\frac{\dot \phi^JP_{,X^{IJ}}Q^I}{H}\left(P_{,X^{LM}}X_1^{LM}
+P_{,K}Q^K\right) +2P_{,X^{IJ}K}Q^K
\left(-\dot{\phi}^I \dot{\phi}^J \alpha_1+\dot
\phi^I\dot Q^J\right)+P_{,KL}Q^KQ^L \nonumber\\
&&+P_{,X^{IJ}}\left(3 \dot{\phi}^I \dot{\phi}^J\alpha_1^2-
4\alpha_1\dot\phi^I\dot Q^J\right)\Bigg]\,,
\end{eqnarray}
where
\begin{equation}
X_1^{IJ}\equiv-\alpha_1 \dot{\phi}^I \dot{\phi}^J
+\dot\phi^{(I}\dot Q^{J)}\,.
\end{equation}
After integrating by parts in the action and employing
the background field equations, the second order action
can be finally written in the rather simple form
\begin{eqnarray}
S_{(2)} &=& \frac12 \int dt d^3 x a^3
\bigl[ (P_{,X^{IJ}} + P_{,X^{IK} X^{JL}} \dot{\phi}^K
\dot{\phi}^L ) \dot{Q}^I \dot{Q}^J
\nonumber\\
&& - \frac{1}{a^2} P_{,X^{IJ}}
\partial_i Q^I \partial^i Q^J
-{\cal{M}}_{IJ} Q^I Q^J + {\cal N}_{IJ}
Q^J \dot{Q}^I \bigr]
\,,
\label{2nd_order_action}
\end{eqnarray}
with the effective squared mass matrix
\begin{eqnarray}
{\cal{M}}_{IJ} &=& -P_{,IJ} + \frac{X^{LM}}{H}\dot{\phi}^K
(P_{,X^{JK}} P_{,I X^{LM}}+P_{,X^{IK}} P_{,J X^{LM}})
\nonumber\\
&&-\frac{1}{H^2} X^{MN} X^{PQ} P_{,X^{MN} X^{PQ}}
P_{,X^{IK}} P_{,X^{JL}} \dot{\phi}^K \dot{\phi}^L
\nonumber\\
&&-\frac{1}{a^3} \frac{d}{dt}
\left[\frac{a^3}{H} P_{,X^{IK}} P_{,X^{JL}}
\dot{\phi}^K \dot{\phi}^L\right]
\,,\\
{\cal{N}}_{IJ} &=&
2\left(P_{,J X^{IK}}-
\frac{X^{MN}}{H} P_{,X^{IK} X^{MN}} P_{,X^{JL}}
\dot{\phi}^L\right) \dot{\phi}^K\,.
\end{eqnarray}
In the same way, the third order action is given by
\begin{eqnarray}
S_{(3)}&=&\int
dtd^3xa^3\Bigg[\left[3H^2\alpha_1^2+\frac{2H}{a^2}\alpha_1\partial^2\psi_1+\frac{1}{2a^4}
\left(\partial^2\psi_1\partial^2\psi_1-\partial_i\partial_j\psi_1\partial^i\partial^j\psi_1\right)\right]\alpha_1
\nonumber
\\
&&+\bigg[-\frac12 \alpha_1^3
\dot{\phi}^I \dot{\phi}^J +\alpha_1^2\dot \phi^I\dot Q^J
+a^{-2}\alpha_1\dot \phi^I\partial^i\psi_1\partial_i
Q^J-\frac{\alpha_1}{2}\dot Q^I \dot
Q^J-a^{-2}\partial_iQ^J\left(\dot
Q^I\partial^i\psi_1+\frac{1}{2}\alpha_1\partial^iQ^I\right)
\bigg]P_{,X^{IJ}}\nonumber
\\
&&+\Big[\alpha_1^2 \dot{\phi}^I \dot{\phi}^J-
\frac{3}{2}\alpha_1\dot \phi^I\dot
Q^J+\frac{1}{2}\dot Q^I \dot Q^J -a^{-2}\partial_iQ^J\left(\dot
\phi^I\partial^i\psi_1+\frac{1}{2}\partial^iQ^I\right)\Big]X_1^{LM}P_{,X^{IJ}X^{LM}}
\nonumber
\\
&&+\Big[\frac12 \alpha_1^2
\dot{\phi}^I \dot{\phi}^J -\alpha_1\dot \phi^I\dot
Q^J+\frac{1}{2}\dot Q^I \dot Q^J -a^{-2}\partial_iQ^J\left(\dot
\phi^I\partial^i\psi_1+\frac{1}{2}\partial^iQ^I\right)\Big]P_{,X^{IJ}K}Q^K\nonumber
\\
&&+\frac{\alpha_1}{2}P_{,IJ}Q^IQ^J+\frac{1}{6}X_1^{IJ}X_1^{LM}X_1^{QR}P_{,X^{IJ}X^{LM}X^{QR}}+\frac{1}{2}X_1^{IJ}X_1^{LM}P_{,X^{IJ}X^{LM}K}Q^K
\nonumber
\\
&&+\frac{1}{2}X_1^{IJ}P_{,X^{IJ}LM}Q^LQ^M+\frac{1}{6}P_{,IJK}Q^IQ^JQ^K\Bigg]\,.
\end{eqnarray}

%%%%%%%%%%%%%%%%%%%%%%%%%%%%%%%%%%%%%%%%%%%%%%%%%%%%%%%%%%%%%%%%%%%%%%%%%%%%%%%%%%%%%%%%%%%%%%%%%%%%%%%%%%%%%%%%%%%%%%%%%%%
\subsection{Decomposition into adiabatic and entropy perturbations}
We can decompose the perturbations into the instantaneous
adiabatic and entropy perturbations, where the adiabatic direction
corresponds to the direction of the background fields' evolution
while the entropy directions are orthogonal to this
\cite{Gordon:2000hv}. For this purpose, following
\cite{Langlois:2008mn}, we introduce an orthogonal basis
${e_{n}^I} (n=1,2,...,N)$ in the field space. The orthonormal
condition is defined as
\begin{equation}
P_{,X^{IJ}} e_n^I e_m^J = \delta_{nm}\,,
\label{orthonormal}
\end{equation}
so that the gradient term
$P_{,X^{IJ}} \partial_i Q^I \partial^i
Q^J$ is diagonalized \cite{vierbein_normal}. We pick up the adiabatic vector as
\begin{equation}
e_1^I = \frac{\dot{\phi}^I}{\sqrt{P_{,X^{JK}} \dot{\phi}^J \dot{\phi}^K}}\,,
\end{equation}
which satisfies the normalization given by
Eq.~(\ref{orthonormal}). The field perturbations are decomposed on
this basis as
\begin{equation}
Q^I =Q_n e_{n}^I\,.
\end{equation}
We defined the matrix $Z_{mn}$ which describes the time variation
of the basis as
\begin{equation}
\dot{e}^{I}_n = e^I_m Z_{mn}\,,
\end{equation}
which satisfies
$Z_{mn} = - Z_{nm}- \dot{P}_{,X^{IJ}}  e^I_m e^J_n$ as a consequence of
$(P_{,X^{IJ}} e_n^I e_m^J)^{\dot{}} =0$.

In terms of the decomposed fields, the second order
action (\ref{2nd_order_action}) can be rewritten as
\begin{eqnarray}
S_{(2)} &=& \frac12 \int dt d^3 x a^3
\bigl[ {\cal K}_{mn} (D_t Q_m)(D_t Q_n)
- \frac{1}{a^2} \delta_{mn}
\partial_i Q_m \partial^i Q_n
\nonumber\\
&& -
{\cal{M}}_{mn} Q_m Q_n
+ {\cal{N}}_{mn} Q_n (D_t Q_m)
 \bigr]
\,,
\label{2nd_order_action_dec}
\end{eqnarray}
where
\begin{eqnarray}
D_t Q_m &\equiv& \dot{Q}_m + Z_{mn} Q_n\,,\\
{\cal K}_{mn} &\equiv& \delta_{mn} +
(P_{,X^{MN}} \dot{\phi}^M \dot{\phi}^N)
P_{,X^{IK} X^{JL}} e_1^{I} e_n^{K} e_1^{J} e_m^{L}\,,\\
{\cal{M}}_{mn} &\equiv& {\cal{M}}_{IJ} e^I_m e^J_n\,,\\
{\cal{N}}_{mn} &\equiv& {\cal{N}}_{IJ} e^I_m e^J_n.
\end{eqnarray}
From the constructions, ${\cal{K}}_{mn}$, ${\cal{M}}_{mn}$
and ${\cal{N}}_{mn}$ are symmetric with respect to $m$ and $n$.
The explicit form of the effective squared
mass matrix in this representation is
\begin{eqnarray}
{\cal{M}}_{mn} &=&
-P_{,mn} + \frac1H\
(P_{,X^{KL}} \dot{\phi}^K \dot{\phi}^L )^{3/2}
(P_{,m X^{MN}} e^M_1 e^N_1)\delta_{n1}\nonumber\\
&&-\frac{1}{4 H^2} (P_{,X^{KL}} \dot{\phi}^K \dot{\phi}^L )^3
(P_{,X^{MN} X^{PQ}} e^M_1 e^N_1 e^P_1 e^Q_1)
\delta_{m1} \delta_{n1}\nonumber\\
&& -\frac{1}{a^3} \frac{d}{dt}
\left[ \frac{a^3}{H}
(P_{,X^{MN}} \dot{\phi}^M \dot{\phi}^N)
P_{,X^{IK}} P_{,X^{JL}} e^K_1 e^L_1 \right]
e^I_m e^J_n\,, \\
{\cal{N}}_{mn} &\equiv&
-\frac1H (P_{,X^{PQ}} \dot{\phi}^P \dot{\phi}^Q)^2
(P_{,X^{KL}X^{MN}} e^K_m e^L_1 e^M_1 e^N_1)
\delta_{n1}\nonumber\\
&&+2\sqrt{P_{,X^{LM}} \dot{\phi}^L \dot{\phi}^M}
(P_{,n X^{IK}} e^I_m e^K_1)\,,
\end{eqnarray}
where $P_{,mn} \equiv P_{,IJ} e^I_m e^J_n$
and
$P_{,n X^{IK}} \equiv P_{,J X^{IK}} e^J_n$.

The equation of motion is obtained as
\begin{eqnarray}
&&\frac{1}{a^3} \frac{d}{dt}
\left[a^3 (2 {\cal{K}}_{mr} D_t Q_m + {\cal{N}}_{mr} Q_m)
\right]
-(2 {\cal{K}}_{mn} Z_{nr} + {\cal{N}}_{mr} ) D_t Q_m
\nonumber\\
&&-\left(2 {\cal{M}}_{mr} + {\cal{N}}_{mn} Z_{mr} \right)
Q_m+\frac{2}{a^2} \partial^2 Q_r=0\,.
\end{eqnarray}

%%%%%%%%%%%%%%%%%%%%%%%%%%%%%%%%%%%%%%%%%%%%%%%%%%%%%%%%%%%%%%%%%%%%%%%%%%%%%%%%%%%%%%%%%%%%%%%%%%%%%%%%%%%%%%%%%%%%%%%%%%%
%%%%%%%%%%%%%%%%%%%%%%%%%%%%%%%%%%%%%%%%%%%%%%%%%%%%%%%%%%%%%%%%%%%%%%%%%%%%%%%%%%%%%%%%%%%%%%%%%%%%%%%%%%%%%%%%%%%%%%%%%%%
\section{Linear perturbations}
In this section, we study the linear order perturbations using the
second order action derived in the previous section.

%%%%%%%%%%%%%%%%%%%%%%%%%%%%%%%%%%%%%%%%%%%%%%%%%%%%%%%%%%%%%%%%%%%%%%%%%%%%%%%%%%%%%%%%%%%%%%%%%%%%%%%%%%%%%%%%%%%%%%%%%%%
\subsection{K-inflation}
Let us consider K-inflation models where $P(X^{IJ}, \phi^I)$ is a
function of only the trace $X = X^{IJ} G_{IJ}(\phi^K)$ of the
kinetic terms where $G_{IJ}(\phi^K)$ is a metric in the field
space:
\begin{equation}
P(X^{IJ}, \phi^I) = \tilde{P}(X, \phi^I).
\end{equation}
The derivatives of P can be evaluated as
\begin{eqnarray}
P_{,X^{IJ}} &=&  G_{IJ} \tilde{P}_{,X}\,,\\
P_{,I} &=&
\frac12 G_{JK,I} \dot{\phi}^J \dot{\phi}^K \tilde{P}_{,X}
+ \tilde{P}_{,I}\,,\\
P_{,X^{IJ} X^{KL}} &=& G_{IJ} G_{KL} \tilde{P}_{,XX}\,,\\
P_{,X^{IJ} K} &=&  \frac12 G_{LM,K} \dot{\phi}^L \dot{\phi}^M
G_{IJ} \tilde{P}_{,XX} + G_{IJ,K} \tilde{P}_{,X}
+ G_{IJ} \tilde{P}_{,XK}\,,\\
P_{,IJ} &=&  \frac14 G_{KL,I} G_{MN,J}
\dot{\phi}^K \dot{\phi}^L \dot{\phi}^M \dot{\phi}^N
\tilde{P}_{,XX} +\frac12 G_{KL,IJ} \dot{\phi}^K \dot{\phi}^L
\tilde{P}_{,X}
\nonumber\\
&&+\frac12 \dot{\phi}^M \dot{\phi}^N(G_{MN,J} \tilde{P}_{,XI}
+  G_{MN,I} \tilde{P}_{,XJ} )
+ \tilde{P}_{,IJ}\,,
\end{eqnarray}
and the sound speed is defined as
\begin{eqnarray}
c_s^2 \equiv  \frac{\tilde{P}_{,X}}
{\tilde{P}_{,X} + 2X \tilde{P}_{,XX}}\,.
\end{eqnarray}

In terms of the decomposed field, the second order action
can be written as
\begin{eqnarray}
S_{(2)} &=&
\frac12 \int dt d^3 x a^3
\biggl[ \left\{ \delta_{mn} + \left(\frac{1}{c_s^2} -1\right)
\delta_{1m} \delta_{1n} \right\}
(D_t Q_m) (D_t Q_n)
\nonumber\\
&&- \frac{1}{a^2} \delta_{mn}
\partial_i Q_m \partial^i Q_n
-{\cal{M}}_{mn} Q_m Q_n
+ {\cal{N}}_{mn} Q_n (D_t Q_m)
\biggr]\,,
\label{2nd_order_action_kinf_dec}
\end{eqnarray}
where we do not show the explicit forms of
${\cal{M}}_{mn}$ and ${\cal{N}}_{mn}$.
%\begin{eqnarray}
%{\cal{M}}_{mn} &=&
%-X^2 \tilde{P}_{,X}^2 \tilde{P}_{,XX}
%(G_{KL,I} e^K_1 e^L_1 e^I_m)
%(G_{MN,J} e^M_1 e^N_1 e^J_n)\nonumber\\
%&&-X \tilde{P}_{,X}^2 (G_{KL,IJ} e^K_1 e^L_1 e^I_m e^J_n)
%-\tilde{P}_{,mn}\nonumber\\
%&&-X \tilde{P}_{,X}
%\left[ (G_{MN,J} e^M_1 e^N_1 e^J_n) \tilde{P}_{,Xm}
%+(G_{MN,I} e^M_1 e^N_1 e^J_m) \tilde{P}_{,Xn}\right]\nonumber\\
%&&+\frac{1}{H}\delta_{n1}(2 \tilde{P}_{,X} X)^{\frac32}
%\biggl[
%(G_{PQ,K} e^P_1 e^Q_1 e^K_m) X \tilde{P}_{,XX}
%+ (G_{MN,K} e^M_1 e^N_1 e^K_m) \tilde{P}_{,X}\nonumber\\
%&&
%+\frac{\tilde{P}_{,Xm}}{\tilde{P}_{,X}}\biggr]
%-\frac{2 \tilde{P}_{,X} X^3}{H^2} \tilde{P}_{,XX}
%\delta_{m1} \delta_{n1}\nonumber\\
%&&-\frac{1}{a^3} \frac{d}{dt}
%\left[ \frac{2a^3}{H} X \tilde{P}^2_{,X}
%(\tilde{P}_{,X} + X \tilde{P}_{,XX})
%e_{I1} e_{J1}\right] e^I_m e^J_n\,,\\
%{\cal{N}}_{mn} &=& 2\sqrt{2 X \tilde{P}_{,X}} \biggl[
%X\tilde{P}_{,XX} (G_{LM,J} e^L_1 e^M_1 e^J_n) \delta_{m1}
%\nonumber\\
%&&+\tilde{P}_{,X} (G_{IK,J} e^I_m e^K_1 e^J_n)
%+\frac{\tilde{P}_{,Xn}}{\tilde{P}_{,X}} \delta_{m1}\biggr]\,,
%\end{eqnarray}
%with $\tilde{P}_{,mn} \equiv \tilde{P}_{,IJ} e^I _m e^J_n$,
%$\tilde{P}_{,Xn} \equiv \tilde{P}_{,XI} e^I _n$.

The sound speed agrees with the adiabatic sound speed defined by
$c_s^2 = dP/dE$. The fact that the sound speeds for the entropy
perturbations are unity has been recognized in
Ref.~\cite{Langlois:2008mn}. This is because the non-trivial
second derivative of $P$ only affects the adiabatic perturbations,
which is the consequence of the fact that the entropy field $s$
satisfies $\dot{s}=0$ in the background  and thus it has no first
order perturbations.

%%%%%%%%%%%%%%%%%%%%%%%%%%%%%%%%%%%%%%%%%%%%%%%%%%%%%%%%%%%%%%%%%%%%%%%%%%%%%%%%%%%%%%%%%%%%%%%%%%%%%%%%%%%%%%%%%%%%%%%%%%%
\subsection{DBI-inflation \label{subsec:DBI}}
An interesting class of models is the DBI-inflation
which describes the motion of a D3 brane in a higher dimensional
spacetime. The DBI action is given by
\begin{equation}
S=-\int d^4 x \frac{1}{f(\phi^K)}\sqrt{-\mbox{det} [g_{\mu \nu} +
f(\phi^K) G_{IJ}(\phi^K) \partial_{\mu} \phi^I
\partial_{\nu} \phi^J]}.
\end{equation}
Recently it was pointed out by Ref.~\cite{Langlois:2008wt} that
the multi-field DBI-inflation is {\it not} included in the
multi-field K-inflation discussed in the previous subsection.
Indeed, $P(X^{IJ})$ is not a function of $X$, but it is given by
\begin{equation}
P(X^{IJ}, \phi^I) = \tilde{P}(\tilde{X}, \phi^I), \quad \tilde{X}
= \frac{(1-{\cal D})}{2f}\,,\label{tildePtildeX}
\end{equation}
where
\begin{eqnarray}
{\cal D} &=& \mbox{det}(\delta^I_J - 2 f X^I_J)\nonumber\\
&=& 1- 2 f G_{IJ} X^{IJ} + 4 f^2 X_I^{[I} X_J^{J]} -8 f^3 X_I^{[I}
X_J^{J} X_K^{K]}+ 16 f^4 X_I^{[I} X_J^J X_K^K
X_L^{L]}.\label{determinant}
\end{eqnarray}
In the background, $\tilde{X}=X$. However, this does
not mean that the full action is a function of
$X$ only. The DBI action takes a specific form of $\tilde{P}$
\begin{equation}
\tilde{P}(\tilde{X}, \phi^I) = -\frac{1}{f}\left( \sqrt{1-2f
\tilde{X}}-1 \right)-V(\phi^I),
\end{equation}
where we allow for a potential $V(\phi^I)$. The sound speed is
defined as
\begin{equation}
c_s^2 \equiv \frac{\tilde{P}_{,\tilde{X}}}
{\tilde{P}_{,\tilde{X}}+ 2X \tilde{P}_{,\tilde{X}
\tilde{X}}}\,.\label{DBIcs}
\end{equation}

The derivatives of $P$ can be calculated as
\begin{eqnarray}
P_{,X^{IJ}} &=& \tilde{P}_{,\tilde{X}} \left(
\frac{d \tilde{X}}{ d X^{IJ}} \right)\,, \\
P_{,X^{IJ}X^{KL}} &=&
\tilde{P}_{,\tilde{X} \tilde{X}} \left(
\frac{d \tilde{X}}{ d X^{IJ}} \right)
\left(
\frac{d \tilde{X}}{ d X^{KL}} \right)
+ \tilde{P}_{,\tilde{X}}
\left(\frac{d^2 \tilde{X}}{d X^{IJ} dX^{KL}}
\right)\,,
\end{eqnarray}
where
\begin{eqnarray}
\frac{d\tilde{X}}{ d X^{IJ}} &=& c_s^2 G_{IJ} +2 f X_{IJ}\,,\\
\frac{d^2 \tilde{X}}{d X^{IJ} dX^{KL}}
&=& -2f \left( G_{IJ} G_{KL} - \frac{1}{2}  G_{IK}G_{JL}
-\frac{1}{2}G_{IL}G_{JK} \right) + O(X^{IJ})\,.
\end{eqnarray}
Here we do not explicitly write down the higher order
terms in $X^{IJ}$ in the second derivative as they
will not contribute to the final result.
In the following, we will omit these terms.
We can also show that
\begin{eqnarray}
P_{,I} &=&
\frac12 G_{JK,I} \dot{\phi}^J \dot{\phi}^K
\tilde{P}_{,\tilde{X}}
+ \tilde{P}_{,I}\,,\\
P_{,IJ} &=&  \frac14 G_{KL,I} G_{MN,J}
\dot{\phi}^K \dot{\phi}^L \dot{\phi}^M \dot{\phi}^N
\tilde{P}_{,\tilde{X}\tilde{X}} +
\frac12 G_{KL,IJ} \dot{\phi}^K \dot{\phi}^L
\tilde{P}_{,\tilde{X}}
\nonumber\\
&&+\frac12 \dot{\phi}^M \dot{\phi}^N
(G_{MN,J} \tilde{P}_{,\tilde{X} I}
+  G_{MN,I} \tilde{P}_{,\tilde{X} J} )
+ \tilde{P}_{,IJ}\,,\\
P_{,X^{IJ} K} &=& \bigl((1-2fX)G_{IJ,K} - 2f_{,K} X G_{IJ}
-2f G_{LM,K} X^{LM} G_{IJ}\nonumber\\
&& 2 f_{,K} X_{IJ} + 2f G_{IL,K} X^L _{\;\;J}
+ 2f G_{JM,K} X_{I} ^{\;\;M}\bigr)\tilde{P}_{,\tilde{X}}
\nonumber\\
&&+(c_s^2 G_{IJ} +2fX_{IJ}) \left( \frac12 G_{LM,K} \dot{\phi}^L
\dot{\phi}^M \tilde{P}_{,\tilde{X}\tilde{X}} +
\tilde{P}_{,\tilde{X} K} \right)\,.
\end{eqnarray}
It is worth noting that even though
$P_{,X^{IJ} K}$ seems to be a bit complicated,
we can show that
\begin{eqnarray}
P_{,X^{IJ} K} \dot{\phi}^J =
\frac12 G_{LM,K} \dot{\phi}^L \dot{\phi}^M
\dot{\phi}_I \tilde{P}_{,\tilde{X}\tilde{X}}
+ \dot{\phi}_I \tilde{P}_{,\tilde{X} K}
+G_{IJ,K}\dot{\phi}^J \tilde{P}_{,\tilde{X}}\,,
\end{eqnarray}
which is just the same form as the K-inflation case. We can also
show that
\begin{equation}
P_{,X^{IJ}} \dot{\phi}^I \dot{\phi}^J
=2 X \tilde{P}_{,\tilde{X}}\,.
\end{equation}

The orthonormality conditions for the basis give
\begin{eqnarray}
e_{n}^I e_{m I} &=&
\frac{1}{\tilde{P}_{,\tilde{X}} c_s^2}
\delta_{mn}-\frac{1}{\tilde{P}_{,\tilde{X}}}
\frac{1-c_s^2}{c_s^2} \delta_{m1} \delta_{n1}\,.
\end{eqnarray}
Using these results, the second order action can be written in terms
of the decomposed perturbations as
\begin{eqnarray}
S_{(2)} &=&
\frac12 \int dt d^3 x a^3
\biggl[ \frac{1}{c_s^2} \delta_{mn}(D_t Q_m) (D_t Q_n)
- \frac{1}{a^2} \delta_{mn}
\partial_i Q_m \partial^i Q_n
\nonumber\\
&&-{\cal{M}}_{mn} Q_m Q_n
+ {\cal{N}}_{mn} Q_n (D_t Q_m)
\biggr]\,.
\label{2nd_order_action_DBIinf_dec}
\end{eqnarray}
%where
%\begin{eqnarray}
%{\cal{M}}_{mn} &=&
%-X^2 \tilde{P}_{,\tilde{X}}^2
%\tilde{P}_{,\tilde{X}\tilde{X}}
%(G_{KL,I} e^K_1 e^L_1 e^I_m)
%(G_{MN,J} e^M_1 e^N_1 e^J_n)\nonumber\\
%&&-X \tilde{P}_{,\tilde{X}}^2 (G_{KL,IJ}
%e^K_1 e^L_1 e^I_m e^J_n)
%-\tilde{P}_{,mn}\nonumber\\
%&&-X \tilde{P}_{,\tilde{X}}
%\left[ (G_{MN,J} e^M_1 e^N_1 e^J_n)
%\tilde{P}_{,\tilde{X}m}
%+(G_{MN,I} e^M_1 e^N_1 e^J_m)
%\tilde{P}_{,\tilde{X}n}\right]\nonumber\\
%&&+\frac{1}{H}\delta_{n1}
%(2 \tilde{P}_{,\tilde{X}} X)^{\frac32}
%\biggl[
%(G_{PQ,K} e^P_1 e^Q_1 e^K_m)
%X \tilde{P}_{,\tilde{X}\tilde{X}}
%+ (G_{MN,K} e^M_1 e^N_1 e^K_m) \tilde{P}_{,\tilde{X}}
%\nonumber\\
%&&
%+\frac{\tilde{P}_{,\tilde{X}m}}{\tilde{P}_{,\tilde{X}}}\biggr]
%-\frac{2 \tilde{P}_{,\tilde{X}} X^3}{H^2}
%\tilde{P}_{,\tilde{X}\tilde{X}}
%\delta_{m1} \delta_{n1}\nonumber\\
%&&-\frac{1}{a^3} \frac{d}{dt}
%\left[ \frac{2a^3}{H} X \tilde{P}^2_{,\tilde{X}}
%(\tilde{P}_{,\tilde{X}} + X \tilde{P}_{,\tilde{X}\tilde{X}})
%e_{I1} e_{J1}\right] e^I_m e^J_n\,,\\
%{\cal{N}}_{mn} &=& 2\sqrt{2 X \tilde{P}_{,\tilde{X}}} \biggl[
%X\tilde{P}_{,\tilde{X}\tilde{X}}
%(G_{LM,J} e^L_1 e^M_1 e^J_n) \delta_{m1}
%\nonumber\\
%&&+\tilde{P}_{,\tilde{X}} (G_{IK,J} e^I_m e^K_1 e^J_n)
%+\frac{\tilde{P}_{,\tilde{X}n}}{\tilde{P}_{,\tilde{X}}}
%\delta_{m1}\biggr]\,,
%\end{eqnarray}
%with $\tilde{P}_{,mn} \equiv \tilde{P}_{,IJ} e^I _m e^J_n$,
%$\tilde{P}_{,\tilde{X}n} \equiv \tilde{P}_{,\tilde{X}I} e^I_n$.

Unlike the K-inflation models, all field perturbations have the
same sound speeds as was pointed out by
Ref.~\cite{Langlois:2008wt}. In order to understand the difference
between K-inflation and the DBI-inflation, we will consider a
generalized model where both cases are included.

%%%%%%%%%%%%%%%%%%%%%%%%%%%%%%%%%%%%%%%%%%%%%%%%%%%%%%%%%%%%%%%%%%%%%%%%%%%%%%%%%%%%%%%%%%%%%%%%%%%%%%%%%%%%%%%%%%%%%%%%%%%
\subsection{Generalized case \label{subsec:generalized_model}}
Let us consider models described by
\begin{equation}
P(X^{IJ}, \phi^I) = \tilde{P}(Y, \phi^I)\,,
\end{equation}
where
\begin{equation}
Y=G_{IJ}(\phi^K) X^{IJ} + \frac{b(\phi^K)}{2}
(X^2-X^{J}_I X^I_J)\,.
\end{equation}
The functional form of $Y$ is chosen so that $Y = X \equiv G_{IJ}
X^{IJ}$ in the background as in the DBI-inflation model. This
model includes as particular cases the K-inflation model for $b=0$
and the DBI-inflation for $b=-2f$ and if $\tilde P$ has the DBI
form. This might be surprising as the DBI action contains
additional terms of order $f^2$ and $f^3$ in $\tilde X$ (see
equations (\ref{tildePtildeX}) and (\ref{determinant})), but it
turns out that these terms do not contribute to the second order
action and the leading order third order action.

Following a similar procedure to the previous subsection, the
second order action can be written in terms of the decomposed
perturbations as
\begin{eqnarray}
S_{(2)} &=&
\frac12 \int dt d^3 x a^3
\biggl[ \left\{ \delta_{mn} +
\frac{2X \tilde{P}_{,YY}}{ \tilde{P}_{,Y}}
\delta_{1m} \delta_{1n}
+\frac{bX}{1+bX} (\delta_{n1} \delta_{m1}-\delta_{mn})
\right\}
(D_t Q_m) (D_t Q_n)
\nonumber\\
&&- \frac{1}{a^2} \delta_{mn}
\partial_i Q_m \partial^i Q_n
-{\cal{M}}_{mn} Q_m Q_n + {\cal{N}}_{mn} Q_n (D_t
Q_m) \biggr]\,, \label{2nd_order_action_general_dec}
\end{eqnarray}

Now we are in a position to explain the difference between
K-inflation and DBI-inflation. As in the K-inflation case, the
non-trivial second derivative of $P$ affects only the adiabatic
perturbations. On the other hand, the non-linear terms of $X^{IJ}$
in $Y$ only affects the entropy perturbations as they vanish in
the background. Then the sound speed for adiabatic perturbations
$c_{ad}^{2}$ and for entropy perturbations $c_{en}^{2}$ are given
by
\begin{equation}
c_{ad}^{2} \equiv
\frac{\tilde{P}_{,Y}}{\tilde{P}_{,Y} + 2X \tilde{P}_{,YY}}\,,
\quad c_{en}^{2} \equiv 1 + bX\,,
\end{equation}
and they are independently determined by $\tilde{P}_{,YY}$
and $d^2 Y/
(dX^{IJ} dX^{KL})$ respectively. Thus in general they are
different. Let us derive the condition under which the two sound
speeds are the same, i.e., $c_{ad}^{2}=c_{en}^{2}$. This condition
is given by
\begin{equation}
2X \frac{\tilde{P}_{,YY}}{\tilde{P}_{,Y}}
 = -\frac{bX}{1+bX}.
\end{equation}
Then we find that the DBI action is a solution of
this equation where $b=-2f$ \cite{GT}.

%%%%%%%%%%%%%%%%%%%%%%%%%%%%%%%%%%%%%%%%%%%%%%%%%%%%%%%%%%%%%%%%%%%%%%%%%%%%%%%%%%%%%%%%%%%%%%%%%%%%%%%%%%%%%%%%%%%%%%%%
%%%%%%%%%%%%%%%%%%%%%%%%%%%%%%%%%%%%%%%%%%%%%%%%%%%%%%%%%%%%%%%%%%%%%%%%%%%%%%%%%%%%%%%%%%%%%%%%%%%%%%%%%%%%%%%%%%%%%%%%
\section{The leading order in slow-roll three point function}
In this section, we will calculate the leading order in slow-roll
third order action for the generalized model of the previous
subsection and then we shall calculate the leading order three
point function for both adiabatic and entropy directions. Finally
we will obtain the three point function of the comoving curvature
perturbation.

%%%%%%%%%%%%%%%%%%%%%%%%%%%%%%%%%%%%%%%%%%%%%%%%%%%%%%%%%%%%%%%%%%%%%%%%%%%%%%%%%%%%%%%%%%%%%%%%%%%%%%%%%%%%%%%%%%%%%%%%
\subsection{Approximations: slow-roll}
In order to control the calculations and to obtain analytical
results we need to make use of some approximations. We will use
the slow-roll approximation, where we define a set of parameters
and assume that these parameters are always small until the end of
inflation. We define the slow-roll parameters as

\begin{equation}
\epsilon \equiv -\frac{\dot H}{H^2}=\frac{X\tilde P_{,Y}}{H^2},\quad
\eta \equiv \frac{\dot \epsilon}{\epsilon H},
\end{equation}
\begin{equation}
\chi_{ad} \equiv \frac{\dot c_{ad}}{c_{ad}H},\quad
\chi_{en} \equiv \frac{\dot
c_{en}}{c_{en}H}.
\end{equation}
It is important to note that these slow-roll parameters are more
general than the usual slow-roll parameters and that their
smallness does not necessarily imply that the fields are rolling
slowly. Assuming that the parameters $\chi_{ad}$ and $\chi_{en}$
are small implies that the rates of change of the adiabatic and
entropy sound speeds are small, but the sound speeds themselves
can have any value between zero and one.

It is convenient to define a parameter that describes the
non-linear dependence of the lagrangian on the kinetic term as
\begin{equation}
\lambda \equiv \frac{2}{3}X^3\tilde P_{,YYY}+X^2\tilde P_{,YY}.
\end{equation}
We will also assume that the rate of change of this new parameter
is small, as given by
\begin{equation}
l \equiv \frac{\dot \lambda}{\lambda H}.
\end{equation}
At the end of this section, we will show that the size of the
leading order three point function of the fields is fully
determined by five parameters evaluated at horizon crossing:
$\epsilon$, $\lambda$, $H$ and both sound speeds.

It turns out that the equations of motion for both adiabatic and
entropy perturbations at first order form a coupled system of
second order linear differential equations, see Appendix for
details. In general, the coupling (denoted by $\xi$ in equation
(\ref{coupling})) between adiabatic and entropy modes cannot be
neglected but in this work we will study the simpler decoupled
case, where we assume that $\xi$ is small when the scales of
interest cross outside the sound horizons, i.e., we will assume
that $\xi\sim\mathcal{O}(\epsilon)$. With these approximations the
adiabatic and entropy modes are decoupled and the system of
equations of motion can be solved analytically. For simplicity, we
will also assume that the mass term present in the entropy
equation of motion is small, i.e., $\mu_s^2/H^2\ll 1$ (refer to
Appendix for more details). When calculating the leading order
three point functions, we assume that the quantities related to
the time derivatives of the basis vectors given by $Z_{mn}$ are
also slow-roll suppressed. Finally, the calculation of the three
point functions in the next subsections is valid in the limit of
small sound speeds. Our results will also include sub-leading
terms of $\mathcal{O}(1)$ but these terms will in general (for
small sound speeds) receive corrections coming from terms of the
order of $\epsilon/c_s^2$, that we have neglected.

%%%%%%%%%%%%%%%%%%%%%%%%%%%%%%%%%%%%%%%%%%%%%%%%%%%%%%%%%%%%%%%%%%%%%%%%%%%%%%%%%%%%%%%%%%%%%%%%%%%%%%%%%%%%%%%%%%%%%%%%
\subsection{Third order action at leading order}
At leading order in the previous approximations, the third order
action for the general model (\ref{action}) is calculated as
\begin{eqnarray}
S_{(3)} &=& \frac{1}{2} \int dx^3 dt a^3 \left[ P_{,X^{IK}X^{JL}}
\dot{\phi}^{(I} \dot{Q}^{K)} \dot{Q}^{J} \dot{Q}^L +\frac{1}{3}
P_{,X^{IK}X^{JL}X^{MN}} \dot{\phi}^{(I} \dot{Q}^{K)}
\dot{\phi}^{(J} \dot{Q}^{L)} \dot{\phi}^{(M} \dot{Q}^{N)}
\right. \nonumber\\
&& \left. -\frac{1}{a^2} P_{,X^{IK}X^{JL}}
\dot{\phi}^{(I} \dot{Q}^{K)}
\partial_i Q^J \partial^i Q^L \right]\,.
\end{eqnarray}
After decomposition into the new adiabatic/entropy basis the third
order action can be written as
\begin{equation}
S_{(3)} =\int dx^3 dt a^3 \left[\frac{1}{2} \Xi_{nml} \dot Q_n
\dot Q_m \dot Q_l -\frac{1}{2a^2} \Upsilon_{nml} \dot
Q_n(\partial_i Q_m)(\partial^i Q_l)
 \right]\,,\label{leadingorderaction}
\end{equation}
where we define the coefficients $\Xi_{nml}$ and $\Upsilon_{nml}$
as
\begin{eqnarray}
\Xi_{nml} &=&
P_{,X^{IK}X^{JL}} \sqrt{P_{,X^{MN}} \dot{\phi}^M \dot{\phi}^N}
e_1^{(I} e_{(n}^{K)} e_m^J e_{l)}^L \nonumber\\
&& +\frac{1}{3} P_{,X^{IK}X^{JL}X^{MN}}
(P_{,X^{PQ}} \dot{\phi}^P \dot{\phi}^Q)^{3/2}
e_1^{(I}e_n^{K)} e_1^{(J} e_m^{L)} e_1^{(M} e_{l}^{N)}\,,\\
\Upsilon_{nml} &=& P_{,X^{IK}X^{JL}} \sqrt{P_{,X^{MN}}
\dot{\phi}^M \dot{\phi}^N}
e_1^{(I} e_n^{K)} e_m^J e_l^L\,.
\end{eqnarray}
We shall now give some useful formulae of the previous quantities
for the different inflationary models considered in this work.

\subsubsection{K-inflation}
For the K-inflation model we have
\begin{eqnarray}
\Xi_{nml} &=&
(2X \tilde{P}_{,X})^{-\frac{1}{2}}
\left(
\frac{2X \tilde{P}_{,XX}}{\tilde{P}_{,X}}
\delta_{1 (n} \delta_{ml)}
+ \frac{4}{3}\frac{X^2 \tilde{P}_{,XXX}}{\tilde{P}_{,X}}
\delta_{n1}\delta_{m1}\delta_{l1}
\right)
, \\
\Upsilon_{nml} &=&(2X \tilde{P}_{,X})^{-\frac{1}{2}}
\frac{2X \tilde{P}_{,XX}}{\tilde{P}_{,X}} \delta_{n1} \delta_{m l}\,.
\end{eqnarray}

\subsubsection{DBI-inflation}
For the DBI-inflation scenario they are given by
\begin{eqnarray}
\Xi_{nml} &=&
(2X \tilde{P}_{,\tilde{X}})^{-\frac{1}{2}}
\frac{1-c_s^2}{c_s^4}
\delta_{1 (n} \delta_{ml)}\,, \\
\Upsilon_{nml} &=& (2X \tilde{P}_{,\tilde{X}})^{-\frac{1}{2}}
\left( \frac{1-c_s^2}{c_s^2} \delta_{n1} \delta_{m l} -2
\frac{1-c_s^2}{c_s^2} \left( \delta_{n1} \delta_{ml} - \delta_{n
(m} \delta_{l)1} \right) \right)\,,
\end{eqnarray}
where $c_s^2$ should be understood as the sound speed defined in
Eq.~(\ref{DBIcs}).

\subsubsection{Generalized case}
For the generalized case of subsection
\ref{subsec:generalized_model},  $\Xi_{nml}$ and $\Upsilon_{nml}$
can be written as
%\begin{eqnarray}
%\Xi_{nml} &=&
%(2X \tilde{P}_{,Y})^{-\frac{1}{2}} \left[
%\frac{2X \tilde{P}_{,YY}}{\tilde{P}_{,Y}}
%\left( \frac{1}{1+bX} \right)
%\delta_{1 (n} \delta_{ml)} \right.\nonumber\\
%&& \left. +\left(
%\frac{4}{3}\frac{X^2 \tilde{P}_{,YYY}}{\tilde{P}_{,Y}}
%+ \frac{2 X \tilde{P}_{,YY}}{\tilde{P}_{,Y}} \frac{bX}{1+bX}
%   \right) \delta_{n1}\delta_{m1}\delta_{l1} \right]
%\,, \\
%\Upsilon_{nml} &=&
%(2X \tilde{P}_{,Y})^{-\frac{1}{2}}
%\left(
%\frac{2X \tilde{P}_{,YY}}{\tilde{P}_{,Y}}
%\delta_{n1} \delta_{m l}
%+ \frac{2 b X }{1+bX}
%\left( \delta_{n1} \delta_{ml} -
%\delta_{n (m} \delta_{l)1}
%\right) \right)\,.
%\end{eqnarray}
\begin{eqnarray}
\Xi_{nml} &=&
(2X \tilde{P}_{,Y})^{-\frac{1}{2}} \left[
\frac{(1-c_{ad}^2)}{c_{ad}^2 c_{en}^2}
\delta_{1 (n} \delta_{ml)} +\left(
\frac{4}{3}\frac{X^2 \tilde{P}_{,YYY}}{\tilde{P}_{,Y}}
- \frac{(1-c_{ad}^2)(1- c_{en}^2)}{c_{ad}^2 c_{en}^2}
\right) \delta_{n1}\delta_{m1}\delta_{l1} \right]
\,, \\
\Upsilon_{nml} &=&
(2X \tilde{P}_{,Y})^{-\frac{1}{2}}
\left(
\frac{1-c_{ad}^2}{c_{ad}^2}
\delta_{n1} \delta_{m l}
- \frac{2 ( 1- c_{en}^2) }{c_{en}^2}
\left( \delta_{n1} \delta_{ml} -
\delta_{n (m} \delta_{l)1}
\right) \right)\,,
\end{eqnarray}
and it is obvious that the DBI-inflation is a specific case of the
general model with $c_{ad}^2 = c_{en}^2 = c_s^2$.

%%%%%%%%%%%%%%%%%%%%%%%%%%%%%%%%%%%%%%%%%%%%%%%%%%%%%%%%%%%%%%%%%%%%%%%%%%%%%%%%%%%%%%%%%%%%%%%%%%%%%%%%%%%%%%%%%%%%%%%%
\subsection{The three point functions of the fields}
In this subsection, we derive the three point functions of the
adiabatic and entropy fields in the generalized case and at
leading order in slow-roll and in the small sound speeds limit. We
consider the two-field case with the adiabatic field $\sigma$ and
the entropy field $s$.

The perturbations are promoted to quantum operators like
\begin{equation}
Q_n(\tau,\mathbf{x})=\frac{1}{(2\pi)^3}\int
d^3\mathbf{k}Q_n(\tau,\mathbf{k})e^{i\mathbf{k}\cdot\mathbf{x}},
\end{equation}
where
\begin{equation}
Q_n(\tau,\mathbf{k})=u_n(\tau,\mathbf{k})a_n(\mathbf{k})+u^*_n(\tau,-\mathbf{k})a^\dag_n(-\mathbf{k}).
\end{equation}
$a_n(\mathbf{k})$ and $a^\dag_n(-\mathbf{k})$ are the annihilation
and creation operator respectively, that satisfy the usual
commutation relations
\begin{equation}
\left[a_n(\mathbf{k_1}),a^\dag_m(\mathbf{k_2})\right]=(2\pi)^3\delta^{(3)}(\mathbf{k_1}-\mathbf{k_2})\delta_{nm},
\quad
\left[a_n(\mathbf{k_1}),a_m(\mathbf{k_2})\right]=\left[a_n^\dag(\mathbf{k_1}),a_m^\dag(\mathbf{k_2})\right]=0.
\end{equation}
At leading order the solution for the mode functions is (see
Appendix for details)
\begin{equation}
u_n(\tau,\mathbf{k})=A_n\frac{1}{k^{3/2}}\left(1+ikc_n\tau\right)e^{-ikc_n\tau},
\end{equation}
where $c_n$ stands for either the adiabatic or the entropy sound
speeds.

The two point correlation function is
\begin{equation}
\langle
0|Q_n(\tau=0,\mathbf{k_1})Q_m(\tau=0,\mathbf{k_2})|0\rangle=(2\pi)^3\delta^{(3)}(\mathbf{k_1}+\mathbf{k_2})\mathcal{P}_{Q_n}\frac{2\pi^2}{k_1^3}\delta_{nm},
\end{equation}
where the power spectrum is defined as
\begin{equation}
\mathcal{P}_{Q_n}=\frac{|A_n|^2}{2\pi^2}, \quad
|A_\sigma|^2=\frac{H^2}{2c_{ad}}, \quad
|A_s|^2=\frac{H^2}{2c_{en}},
\end{equation}
and it should be evaluated at the time of horizon crossing
${c_n}_* k_1=a_*H_*$

The vacuum expectation value of the three point operator in the
interaction picture (at first order) is
\cite{Maldacena:2002vr,Weinberg:2005vy}
\begin{equation}
\langle\Omega|Q_l(t,\mathbf{k_1})Q_m(t,\mathbf{k_2})Q_n(t,\mathbf{k_3})|\Omega\rangle=-i\int^t_{t_0}d\tilde
t \langle 0
|\left[Q_l(t,\mathbf{k_1})Q_m(t,\mathbf{k_2})Q_n(t,\mathbf{k_3}),H_I(\tilde
t)\right]|0\rangle,
\end{equation}
where $t_0$ is some early time during inflation when the field's
vacuum fluctuation are deep inside the horizons, $t$ is some time
after horizon exit. $|\Omega\rangle$ is the interacting vacuum
which is different from the free theory vacuum $|0\rangle$. If one
uses conformal time, it's a good approximation to perform the
integration from $-\infty$ to $0$ because $\tau\approx-(aH)^{-1}$.
$H_I$ denotes the interaction hamiltonian and it is given by
$H_I=-L_3$, where $L_3$ is the lagrangian obtained from the action
(\ref{leadingorderaction}).

At this order, the only non-zero three point functions are
\begin{eqnarray}
\langle\Omega|Q_\sigma(0,\mathbf{k_1})Q_\sigma(0,\mathbf{k_2})Q_\sigma(0,\mathbf{k_3})|\Omega\rangle&=&
(2\pi)^3\delta^{(3)}(\mathbf{k_1}+\mathbf{k_2}+\mathbf{k_3})\frac{2c_{ad}|A_\sigma|^6}{H}
\frac{1}{\Pi_{i=1}^3k_i^3}\frac{1}{K} \nonumber\\&&
\times\bigg[6c_{ad}^2(C_3+C_4)\frac{k_1^2k_2^2k_3^2}{K^2}
-C_1k_1^2\mathbf{k_2}\cdot\mathbf{k_3}\left(1+\frac{k_2+k_3}{K}+2\frac{k_2k_3}{K^2}\right)
\nonumber\\
&&+2\,\,\mathrm{cyclic\, terms} \bigg],
\end{eqnarray}
\begin{eqnarray}
\langle\Omega|Q_\sigma(0,\mathbf{k_1})Q_s(0,\mathbf{k_2})Q_s(0,\mathbf{k_3})|\Omega\rangle&=&
(2\pi)^3\delta^{(3)}(\mathbf{k_1}+\mathbf{k_2}+\mathbf{k_3})\frac{|A_\sigma|^2|A_s|^4}{H}
\frac{1}{\Pi_{i=1}^3k_i^3}\frac{1}{\tilde K} \nonumber\\&&
\times\bigg[
C_2c_{en}^2k_3^2\mathbf{k_1}\cdot\mathbf{k_2}\left(1+\frac{c_{ad}k_1+c_{en}k_2}{\tilde
K }+\frac{2c_{ad}c_{en}k_1k_2}{\tilde
K^2}\right)+(k_2\leftrightarrow k_3) \nonumber\\&&
+4C_3c_{ad}^2c_{en}^4\frac{k_1^2k_2^2k_3^2}{\tilde K^2}
-2(C_1+C_2)c_{ad}^2k_1^2\mathbf{k_2}\cdot\mathbf{k_3}\left(1+c_{en}\frac{k_2+k_3}{\tilde
K}+2c_{en}^2\frac{k_2k_3}{\tilde K^2}\right) \bigg],\nonumber\\
\label{tpfmix}
\end{eqnarray}
where $K=k_1+k_2+k_3$, $\tilde K=c_{ad}k_1+c_{en}(k_2+k_3)$,
$\mathrm{cyclic\, terms}$ means cyclic permutations of the three
wave vectors and $(k_2\leftrightarrow k_3)$ denotes a term like
the preceding one but with $k_2$ and $k_3$ interchanged. The pure
adiabatic three point function is evaluated at the moment $\tau_*$
at which the total wave number $K$ exits the horizon, i.e., when
$K{c_{ad}}_*=a_*H_*$. Because of the different propagation speeds,
the adiabatic and entropy modes become classical at different
times, however at leading order we assume that the background
dependent coefficients of (\ref{tpfmix}) do not vary with time and
so they can also be evaluated at the moment $\tau_*$.

The different constants $C_N$ are given by
\begin{eqnarray}
C_1&=&(2H^2\epsilon)^{-\frac{1}{2}} \frac{1-c_{ad}^2}{c_{ad}^2},
\quad C_2=-2(2H^2\epsilon)^{-\frac{1}{2}}
\frac{1-c_{en}^2}{c_{en}^2},
\nonumber\\
C_3&=&(2H^2\epsilon)^{-\frac{1}{2}}
\frac{1-c_{ad}^2}{c_{ad}^2c_{en}^2}, \quad
C_4=(2H^2\epsilon)^{-\frac{1}{2}}\left(
\frac{2\lambda}{H^2\epsilon} -
\frac{1-c_{ad}^2}{c_{ad}^2c_{en}^2}\right).
\end{eqnarray}

%%%%%%%%%%%%%%%%%%%%%%%%%%%%%%%%%%%%%%%%%%%%%%%%%%%%%%%%%%%%%%%%%%%%%%%%%%%%%%%%%%%%%%%%%%%%%%%%%%%%%%%%%%%%%%%%%%%%%%%%
\subsection{The three point function of the comoving curvature perturbation}
In this subsection, we calculate the leading order in slow-roll
three point function of the comoving curvature perturbation in
terms of three point function of the fields obtained in the
previous subsection.

During the inflationary era the comoving curvature perturbation is
given by
\begin{equation}
\mathcal{R}=\frac{H}{\dot\sigma}\frac{Q_\sigma}{\sqrt{\tilde{P}_{,Y}}}.
\end{equation}
It is convenient to define the entropy perturbation $\mathcal{S}$
as
\begin{equation}
S=\frac{H}{\dot\sigma}\frac{Q_s}{\sqrt{\tilde{P}_{,Y}}}\sqrt{\frac{c_{en}}{c_{ad}}},
\end{equation}
so that
$\mathcal{P}_{\mathcal{S}_*}\simeq\mathcal{P}_{\mathcal{R}_*}$,
where the subscript $*$ means that the quantity should be
evaluated at horizon crossing.

In this work we will ignore the possibility that the entropy
perturbations during inflation can lead to primordial entropy
perturbations that could be observable in the CMB. But we shall
consider the effect of entropy perturbations on the final
curvature perturbation. We will follow the analysis of Wands
\emph{et.al.} \cite{Wands:2002bn}, where it has been shown that
even on large scales the curvature perturbation can change in time
because of the presence of entropy perturbations. The way the
entropy perturbations are converted to curvature perturbations is
model dependent but it was shown that this model dependence can be
parameterized by a transfer coefficient $T_{\mathcal{RS}}$
\cite{Wands:2002bn} like
\begin{equation}
\mathcal{R}=\mathcal{R}_*+T_\mathcal{RS}\mathcal{S}_*=\mathcal{A}_\sigma
Q_{\sigma *}+\mathcal{A}_sQ_{s*},
\end{equation}
with
\begin{equation}
\mathcal{A}_\sigma=\left(\frac{H}{\dot\sigma\sqrt{\tilde{P}_{,Y}}}\right)_*,\quad
\mathcal{A}_s=T_\mathcal{RS}\left(\frac{H}{\dot\sigma\sqrt{\tilde{P}_{,Y}}}\sqrt{\frac{c_{en}}{c_{ad}}}\right)_*.
\end{equation}

Using the previous expressions we can now relate the three point
function of the curvature perturbation to the three point
functions of the fields obtained in the previous subsection. The
three point function of the curvature perturbation is given by
\begin{equation}
\langle\mathcal{R}(\mathbf{k_1})\mathcal{R}(\mathbf{k_2})\mathcal{R}(\mathbf{k_3})\rangle=
\mathcal{A}^3_\sigma\langle
Q_\sigma(\mathbf{k_1})Q_\sigma(\mathbf{k_2})Q_\sigma(\mathbf{k_3})\rangle
+ \mathcal{A}_\sigma\mathcal{A}_s^2\left(\langle
Q_\sigma(\mathbf{k_1})Q_s(\mathbf{k_2})Q_s(\mathbf{k_3})\rangle+2\,perms.\right).
\end{equation}

For the DBI-inflation case the previous equation can be simplified
and the total momentum dependence of the three point function of
the comoving curvature perturbation is the same as in single field
DBI \cite{Langlois:2008wt}. For our general model this is no
longer the case, i.e., the different terms of the previous
equation have different momentum dependence. Once again one can
see that DBI-inflation is a very particular case and more
importantly it provides a distinct signature that enables us to
distinguish it from other more general models.

%%%%%%%%%%%%%%%%%%%%%%%%%%%%%%%%%%%%%%%%%%%%%%%%%%%%%%%%%%%%%%%%%%%%%%%%%%%%%%%%%%%%%%%%%%%%%%%%%%%%%%%%%%%%%%%%%%%%%%%%%%%
%%%%%%%%%%%%%%%%%%%%%%%%%%%%%%%%%%%%%%%%%%%%%%%%%%%%%%%%%%%%%%%%%%%%%%%%%%%%%%%%%%%%%%%%%%%%%%%%%%%%%%%%%%%%%%%%%%%%%%%%%%%
\section{Conclusion}
In this paper, we studied the non-gaussianity from the bispectrum
in general multi-field inflation models with a generic kinetic
term. Our model is fairly generic including the K-inflation and
the DBI-inflation as special cases. We derived the second and
third order actions for the perturbations including the effect of
gravity. The second order action is written in terms of adiabatic
and entropy perturbations. It was shown that the sound speeds for
these perturbations are in general different. In the K-inflation
the entropy perturbations propagate at the speed of light. 
The DBI-inflation is a special case where the sound speed for the 
entropy perturbations is the same as the adiabatic sound speed. 
We found that, from the requirement that the sound speeds for 
adiabatic and entropy perturbations are the same, we obtain the DBI 
form for the action.

Then we derive the three point function in the small sound speeds
limit at leading order in slow-roll expansion. In these
approximations there exists a three point function between
adiabatic perturbations $Q_{\sigma}$ and entropy perturbations
$Q_s$, $\langle
Q_{\sigma}(\mathbf{k_1})Q_{s}(\mathbf{k_2})Q_{s}(\mathbf{k_3})
\rangle$, in addition to the pure adiabatic three point function.
This mixed contribution has a different momentum dependence if the
sound speed for the entropy perturbations is different from the
adiabatic one. This provides a possibility to distinguish between
the multi-field models and the single field models. Unfortunately,
in the multi-field DBI case, the sound speed for the entropy
perturbation is the same as the adiabatic one and the mixed
contribution only changes the amplitude of the three point
function. This could help to ease the constraints on DBI-inflation
as is discussed in Ref.~\cite{Langlois:2008wt}.

In order to calculate the effect of the entropy perturbations
on the curvature perturbation at the recombination, we need
to specify a model that describes how the entropy
perturbations are converted to the curvature perturbations.
In addition, even during inflation, if the trajectory in
field space changes non-trivially, the entropy perturbations
can be converted to the curvature perturbation. In this
paper, we assumed that this does not happen and neglected
a mixing between the curvature and entropy perturbations.
It would be interesting to study this mixing in specific
string theory motivated models.

{\it Note added:}
While we were writing up this work, similar results
appeared on the arXiv \cite{Langlois:2008qf}.

%%%%%%%%%%%%%%%%%%%%%%%%%%%%%%%%%%%%%%%%%%%%%%%%%%%%%%%%%%%%%%%%%%%%%%%%%%%%%%%%%%%%%%%%%%%%%%%%%%%%%%%%%%%%%%%%%%%%%%%%%%%
\begin{acknowledgments}
We would like to thank G. Tasinato for useful discussions. FA is
supported by ``Funda\c{c}\~{a}o para a Ci\^{e}ncia e Tecnologia
(Portugal)", with the fellowship's reference number:
SFRH/BD/18116/2004. SM is supported by a JSPS Research Fellowship
and JSPS Core-to-Core Program ``International Research Network for
Dark Energy". SM is grateful to the ICG, Portsmouth, for their
hospitality when this work was initiated. KK is supported by STFC.
\end{acknowledgments}
%%%%%%%%%%%%%%%%%%%%%%%%%%%%%%%%%%%%%%%%%%%%%%%%%%%%%%%%%%%%%%%%%%%%%%%%%%%%%%%%%%%%%%%%%%%%%%%%%%%%%%%%%%%%%%%%%%%%%%%%%%%
%%%%%%%%%%%%%%%%%%%%%%%%%%%%%%%%%%%%%%%%%%%%%%%%%%%%%%%%%%%%%%%%%%%%%%%%%%%%%%%%%%%%%%%%%%%%%%%%%%%%%%%%%%%%%%%%%%%%%%%%%%%
%%%%%%%%%%%%%%%%%%%%%%%%%%%%%%%%%%%%%%%%%%%%%%%%%%%%%%%%%%%%%%%%%%%%%%%%%%%%%%%%%%%%%%%%%%%%%%%%%%%%%%%%%%%%%%%%%%%%%%%%%%%
%%%%%%%%%%%%%%%%%%%%%%%%%%%%%%%%%%%%%%%%%%%%%%%%%%%%%%%%%%%%%%%%%%%%%%%%%%%%%%%%%%%%%%%%%%%%%%%%%%%%%%%%%%%%%%%%%%%%%%%%%%%
%%%%%%%%%%%%%%%%%%%%%%%%%%%%%%%%%%%%%%%%%%%%%%%%%%%%%%%%%%%%%%%%%%%%%%%%%%%%%%%%%%%%%%%%%%%%%%%%%%%%%%%%%%%%%%%%%%%%%%%%%%%
\appendix
\section{Equations of motion for the fluctuations}

Here, we derive the equations of motion for linear perturbations
for the generalized model introduced in
\ref{subsec:generalized_model}. In terms of the field space
``covariant quantities'' \cite{Sasaki:1995aw} which are given by
\begin{eqnarray}
&&{\cal{D}}_t \dot{\phi}^I  \equiv
\ddot{\phi}^I + \Gamma^I _{JK} \dot{\phi}^J  \dot{\phi}^K\,,\\
&&{\cal{D}}_t Q^I  \equiv
\dot{Q}^I + \Gamma^I _{JK} \dot{\phi}^J  Q^K\,,\\
&&{\cal{D}}_I {\cal{D}}_J \tilde{P} \equiv
\tilde{P}_{,IJ} - \Gamma^K _{IJ} \tilde{P}_{,K}\,,\\
&&{\cal{R}}^I _{\;KLJ} \equiv \Gamma^I _{\;KJ,L}-
\Gamma^I _{\;KL,J} + \Gamma^I _{\;LM}\Gamma^M_{\;JK}
-\Gamma^I _{\;JM} \Gamma^M_{\;LK}\,,
\end{eqnarray}
($\Gamma^I _{JK}$ denotes the Christoffel symbols
associated with the field space metric $G_{IJ}$),
the second order action can be expressed as
\begin{eqnarray}
S_{(2)} &=& \frac12 \int dt d^3 x a^3
\biggl[ \left( \tilde{P}_{,Y}  G_{IJ}
+ \tilde{P}_{,YY} \dot{\phi}_I \dot{\phi}_J \right)
{\cal{D}}_t Q^I {\cal{D}}_t Q^J
\nonumber\\
&&- \frac{1}{a^2}  \tilde{P}_{,Y} \left[(1+bX) G_{IJ} - b
X_{IJ}\right]
\partial_i Q^I \partial^i Q^J
-\bar{{\cal{M}}}_{IJ} Q^I Q^J +
2 \tilde{P}_{,YJ} \dot{\phi}_I Q^J {\cal{D}}_t Q^I
\biggr]\,,\nonumber\\
\end{eqnarray}
with the effective squared mass matrix
\begin{eqnarray}
\bar{{\cal{M}}}_{IJ} &=&- {\cal{D}}_I {\cal{D}}_J \tilde{P}
- \tilde{P}_{,Y} {\cal{R}}_{IKLJ} \dot{\phi}^K \dot{\phi}^L
+ \frac{X \tilde{P}_{,Y}}{H} (\tilde{P}_{,YJ} \dot{\phi}_I
+ \tilde{P}_{,YI} \dot{\phi}_J)\nonumber\\
&&+\frac{X \tilde{P}^3 _{,Y}}{2 H^2}(1-\frac{1}{c_{ad}^2})
\dot{\phi}_I \dot{\phi}_J- \frac{1}{a^3} {\cal{D}}_t
\left[\frac{a^3}{2H} \tilde{P}^2 _{,Y}
\left(1+\frac{1}{c_{ad}^2}\right)\dot{\phi}_I \dot{\phi}_J
\right]\,.
\end{eqnarray}

It is worth noting that
except for the coefficients of the kinetic term and
the gradient term, this action is the same
as the K-inflation case and DBI-inflation case
which are derived in \cite{Langlois:2008mn}
and \cite{Langlois:2008wt}, respectively.

From now on we will derive the equations of motion for the
fluctuations. For simplicity, let us now restrict our attention to
the two field case ($I = 1,2$). Then, the perturbations can be
decomposed into $Q^I = Q_\sigma e^I_\sigma + Q_s e^I_s$, where
$e^I_\sigma = e^I_1$ and $e^I_s$ is the unit vector orthogonal to
$e^I_\sigma$. As in standard inflation, it is more convenient to
use conformal time $\tau = \int dt /a(t)$ and define the
canonically normalized fields \cite{mukhanov-sasaki}
\begin{eqnarray}
v_\sigma \equiv \frac{a}{c_{ad}} Q_\sigma\,,\;\;\;\;\;
v_s \equiv \frac{a}{c_{en}}Q_s\,.
\end{eqnarray}

From the similar calculations with K-inflation and DBI-inflation
cases analyzed by \cite{Langlois:2008mn} and
\cite{Langlois:2008wt}, respectively, we find the equations of
motion for $v_\sigma$ and $v_s$ as
\begin{eqnarray}
v_\sigma '' - \xi v_s' + \left( c_{ad}^2 k^2
-\frac{z''}{z} \right) v_\sigma - \frac{(z \xi)'}{z}v_s
&=&0\,,
\label{eom_adiabatic}\\
v_s'' + \xi v_\sigma ' + \left(c_{en}^2 k^2
-\frac{\alpha''}{\alpha} + a^2 \mu_s^2\right) v_s
-\frac{z'}{z} \xi v_\sigma &=& 0
\label{eom_entropy}\,,
\end{eqnarray}
where the primes denote the derivative with respect to $\tau$ and
\begin{eqnarray}
\xi &\equiv& \frac{a}{\dot{\sigma} \tilde{P}_{,Y} c_{ad}}
\left[ (1+ c_{ad}^2) \tilde{P}_{,s} -c_{ad}^2 \dot{\sigma}^2
\tilde{P}_{,Ys}\right]\,,\label{coupling}\\
\mu_s^2 &\equiv&
-\frac{\tilde{P}_{,ss}}{\tilde{P}_{,Y}}
+ \frac12 \dot{\sigma}^2 \tilde{R}
- \frac{1}{2 c_{ad}^2 X}
\frac{\tilde{P}_{,s}^2}{\tilde{P}_{,Y}^2}
+2 \frac{\tilde{P}_{,Ys} \tilde{P}_{,s}}{\tilde{P}_{,Y}^2}\,,\\
z&\equiv& \frac{a \dot{\sigma}}{ c_{ad} H} \sqrt{\tilde{P}_{,Y}}\,,
\;\;\;\;\;\alpha\equiv a \sqrt{\tilde{P}_{,Y}}\,,
\end{eqnarray}
with
\begin{eqnarray}
\dot{\sigma} \equiv \sqrt{2X}\,,\;\;\;\;\;
\tilde{P}_{,s} \equiv \tilde{P}_{,I} e^I_s
\sqrt{\tilde{P}_{,Y}} c_{en}\,,
\;\;\;\;\;\tilde{P}_{,Ys} \equiv \tilde{P}_{,YI}
e^I_s \sqrt{\tilde{P}_{,Y}} c_{en}\,,
\;\;\;\;\;\tilde{P}_{,ss} \equiv
({\cal{D}}_I {\cal{D}}_J \tilde{P})
e^I_s e^J_s \tilde{P}_{,Y}
c_{en}^2 \,,
\end{eqnarray}
and $\tilde{R}$ denotes the Riemann scalar curvature of the
field space.

If we assume that the effect of the coupling $\xi$ can be
neglected when the scales of interest cross the sound horizons the
two degrees of freedom are decoupled and the system can be easily
quantized. If we further assume the slow-roll approximations, the
time evolution of $H$, $c_{ad}$, and $\dot{\sigma}$ is small with
respect to that of the scale factor and the relations $z''/z \simeq
2/\tau^2$ and $\alpha''/\alpha \simeq 2/\tau^2$ hold (see section
V.A for these approximations). The solutions of
(\ref{eom_adiabatic}) and (\ref{eom_entropy}) with the
Bunch-Davies vacuum initial conditions are thus given by
\begin{eqnarray}
v_{\sigma k} &\simeq& \frac{1}{\sqrt{2 k c_{ad}}}
e^{-i k c_{ad} \tau} \left( 1-\frac{i}{k c_{ad} \tau}\right)\,,
\\
v_{s k} &\simeq& \frac{1}{\sqrt{ 2 k c_{en}}}
e^{-i k c_{en} \tau} \left(1-\frac{i}{k c_{en} \tau}\right)\,,
\end{eqnarray}
when $\mu_s^2/H^2$ is negligible for the entropy mode.

Therefore, the power spectra for $Q_\sigma$ and $Q_s$
are obtained as
\begin{eqnarray}
{\cal{P}}_{Q_\sigma} \simeq \frac{H^2}{4 \pi^2 c_{ad}}\,,
\;\;\;\;\;
{\cal{P}}_{Q_s} \simeq \frac{H^2}{4 \pi^2 c_{en}}\,,
\end{eqnarray}
which are evaluated at sound horizon crossing.
The ratio of the power spectra
for the adiabatic and entropy modes is thus given by
${\cal{P}}_{Q_s}/{\cal{P}}_{Q_{\sigma}} = c_{ad}/c_{en}$.

%%%%%%%%%%%%%%%%%%%%%%%%%%%%%%%%%%%%%%%%%%%%%%%%%%%%%%%%%%%%%%%%%%%%%%%%%%%%%%%%%%%%%%%%%%%%%%%%%%%%%%%%%%%%%%%%%%%%%%%%%%%
%%%%%%%%%%%%%%%%%%%%%%%%%%%%%%%%%%%%%%%%%%%%%%%%%%%%%%%%%%%%%%%%%%%%%%%%%%%%%%%%%%%%%%%%%%%%%%%%%%%%%%%%%%%%%%%%%%%%%%%%%%%
%%%%%%%%%%%%%%%%%%%%%%%%%%%%%%%%%%%%%%%%%%%%%%%%%%%%%%%%%%%%%%%%%%%%%%%%%%%%%%%%%%%%%%%%%%%%%%%%%%%%%%%%%%%%%%%%%%%%%%%%%%%
%%%%%%%%%%%%%%%%%%%%%%%%%%%%%%%%%%%%%%%%%%%%%%%%%%%%%%%%%%%%%%%%%%%%%%%%%%%%%%%%%%%%%%%%%%%%%%%%%%%%%%%%%%%%%%%%%%%%%%%%%%%
%%%%%%%%%%%%%%%%%%%%%%%%%%%%%%%%%%%%%%%%%%%%%%%%%%%%%%%%%%%%%%%%%%%%%%%%%%%%%%%%%%%%%%%%%%%%%%%%%%%%%%%%%%%%%%%%%%%%%%%%%%%
%%%%%%%%%%%%%%%%%%%%%%%%%%%%%%%%%%%%%%%%%%%%%%%%%%%%%%%%%%%%%%%%%%%%%%%%%%%%%%%%%%%%%%%%%%%%%%%%%%%%%%%%%%%%%%%%%%%%%%%%%%%
%%%%%%%%%%%%%%%%%%%%%%%%%%%%%%%%%%%%%%%%%%%%%%%%%%%%%%%%%%%%%%%%%%%%%%%%%%%%%%%%%%%%%%%%%%%%%%%%%%%%%%%%%%%%%%%%%%%%%%%%%%%
%%%%%%%%%%%%%%%%%%%%%%%%%%%%%%%%%%%%%%%%%%%%%%%%%%%%%%%%%%%%%%%%%%%%%%%%%%%%%%%%%%%%%%%%%%%%%%%%%%%%%%%%%%%%%%%%%%%%%%%%%%%
%%%%%%%%%%%%%%%%%%%%%%%%%%%%%%%%%%%%%%%%%%%%%%%%%%%%%%%%%%%%%%%%%%%%%%%%%%%%%%%%%%%%%%%%%%%%%%%%%%%%%%%%%%%%%%%%%%%%%%%%%%%
%%%%%%%%%%%%%%%%%%%%%%%%%%%%%%%%%%%%%%%%%%%%%%%%%%%%%%%%%%%%%%%%%%%%%%%%%%%%%%%%%%%%%%%%%%%%%%%%%%%%%%%%%%%%%%%%%%%%%%%%%%%


\begin{thebibliography}{99}
\bibitem{derivative}
Strictly speaking, we adopt the symmetrised derivative,
$P_{,X^{IJ}} \equiv \frac12
\left(\frac{\partial P}{\partial X^{IJ}}
+ \frac{\partial P}{\partial X^{JI}}\right)$.
\end{thebibliography}
\end{document}